\documentclass[prb,twocolumn,superscriptaddress,floatfix,showpacs]{revtex4}
\usepackage[dvips]{graphicx}
\usepackage{dcolumn}
\usepackage{amsmath}
\usepackage{amssymb}
\usepackage{amsbsy}
\begin{document}
\input{epsf}
\title{Effect of the surface polarization in polar perovskites 
 studied from first principles}
\author{M. Fechner}
\affiliation{Max-Planck-Institut f\"ur Mikrostrukturphysik, Weinberg
  2, D-06120 Halle (Saale), Germany}
  \author{S. Ostanin}
\affiliation{Max-Planck-Institut f\"ur Mikrostrukturphysik, Weinberg
  2, D-06120 Halle (Saale), Germany}
\author{I. Mertig}
\affiliation{Max-Planck-Institut f\"ur Mikrostrukturphysik, Weinberg
  2, D-06120 Halle (Saale), Germany}
\affiliation{Martin-Luther-Universit\"at Halle-Wittenberg,
 Fachbereich Physik, D-06099 Halle, Germany}
\date{\today}

\begin{abstract}
The (001) surfaces of polar perovskites BaTiO$_3$ and PbTiO$_3$ have been studied from first principles at T=0~K. For both cases of polarization, the most stable TiO-terminated interfaces show intrinsic ferroelectricity. In the topmost layer, where the O atoms are $>$0.1 \AA~ above Ti, this leads to metallic instead of the insulating behavior of the electronic states that may have important implications for multiferroic tunneling junctions. 
\end{abstract}


\maketitle

Epitaxial growth technique opens the way to combine ferroelectrics (FE) and ferromagnets (FM) into multicomponent multiferroics, which are inaccessible by traditional synthesis.\cite{Dawber,Zheng}  Currently, many laboratories across the world work on a prototypic device, consisting of a few-nm thick ferroelectric sandwiched between a ferromagnet and another metallic contact. External electric field applied to the FE phase causes a switching of its polarization $\bf P$ that, in turn, through the magnetoelectric coupling may change the magnetic order in the ferromagnetic phase.\cite{Eerenstein}  Altering $\bf P$ and magnetization independently for encoding information in multiferroics, the smallest quaternary logical memory might be obtained. Fundamentally new multiferroics require a better understanding of the FE order parameters, especially, in the case of thin films here the symmetry is reduced. 

Since direct measurements of atomic displacements, occurring in FE near the interface, is extremely challenging, their structures can be understood and numerically characterized from first principles. Recently, much work has been conducted to study bare FE surfaces using the {\it ab initio} density functional theory (DFT) calculations.\cite{Padilla1997,CohenBatio31997,S'aghi-Szab'o1998,Heifets2001,Neaton2003,umeno:174111,wang2006,Lisenkov2007,Eglitis2007} It has been found that the critical thickness down to 3 unit cells (1.2 nanometers) is enough to enable the existence of ferroelectricity at room temperature.\cite{Junquera,Spaldin2004,Fong2004} However, there are only some convincing evidences in literature i.e. the work by Cohen \cite{CohenBatio31997} that the direction of $\bf P$ may affect the surface relaxation. The functionality of multiferroics assumes that $\bf P$ must be reversible and parallel to the surface normal. Hence, it is worthwhile to carry out {\it ab initio} calculations which model the reaction of the (001) surface of polar ferroelectric surfaces upon the change of its reversible $\bf P$.  

In this report, we study the (001) surface of ABO$_3$ perovskites (A = Sr, Ba, Pb and B = Ti), which represent a wide class of ferroelectrics ranging from paraelectric SrTiO$_3$ (STO) to highly polar PbTiO$_3$ (PTO), while BaTiO$_3$ (BTO), with its moderate spontaneous polarization $P_s$, is an example of a typical FE. The study is based on extensive calculations, using the Vienna Ab-initio Simulation Package \cite{Kresse96} (VASP), in which the effects of relaxation of atomic positions are included. Nowadays, many FE  properties can be successfully calculated from first principles.\cite{Ghosez}
\begin{table}[b]
\caption{\label{bulk_zellen} The lattice parameters $a$, $c/a$, fractional atomic coordinates, $z/a$, and polarization, calculated for the room temperature PTO, BTO and STO phases, are shown in comparison with the corresponding experimental data.}
\begin{ruledtabular}
\begin{tabular}{lllll}
& & PTO & BTO & STO \\
\hline
a (\AA)& exp. &3.892 $^a$&3.991 $^b$&3.898 $^b$   \\
&&3.858&3.943&3.885\\
c/a& exp.&1.071 $^a$&1.011 $^b$&1.0\\
& & 1.071&1.013      &        \\
\hline
Ti&exp.&0.542 $^a$&   0.489 $^b$      & 0.5 \\
&  &   0.542      &       0.492     & \\
O-1,2& exp. & 0.629 $^a$ &  0.511 $^b$     & 0.5 \\
& & 0.622          &     0.513     & \\
O-3& exp. &0.124 $^a$    &   0.018 $^b$     & 0.0\\
& &   0.115       &    0.021     &    \\
\hline
$P_s$~($\mu C/cm^2$)& exp. &75 $^c$&  26 $^d$    & -\\
& &   94.3        &   22.9     &    \\
\end{tabular}
\end{ruledtabular}
$^a${~Ref. [\onlinecite{Mestric2005}]}, $^b${~Ref. [\onlinecite{Toshio1981}]}, $^c${~Ref. [\onlinecite{Gavrilyachenko1970}]}, $^d${~Ref. [\onlinecite{Wieder1955}]}
\end{table} 
Table \ref{bulk_zellen} collects the experimental data for the lattice parameters and atomic positions, obtained for the room-temperature tetragonal phase of PTO and BTO (with space group symmetry $P4mm$) and for cubic STO ($Pm-3m$), in comparison with our calculation. Overall, there is a good agreement between the measured and theoretical structure parameters for the three systems.  For polar BTO and PTO, their values  of $\bf P$, calculated by the Berry phase approach \cite{King-Smith1993,Resta1994}, are in reasonably good agreement with experiment. The minor differences, seen in Table \ref{bulk_zellen} between our and some other recent DFT results,\cite{S'aghi-Szab'o1998,King-Smith1994,Eglitis2007}  can be attributed to the choice of pseudopotentials and/or to the approximation used for the exchange and correlation potential. We used the local density approximation (LDA) while the electron-ion interactions were described by the PAW pseudopotentials. After relaxation the calculated forces are always less than $0.5\cdot 10^{-2}$eV/\AA. The electron pseudo-wave-functions were represented using plane waves, with a cutoff energy of 650 eV. For the Brillouin-zone integration a dense Monkhorst-Pack \cite{Monkhorst1976} mesh was used. 

We calculated 5-unit-cell ($\sim$2-nm) thick ABO$_3$ film. The atoms of the two upper or, alternatively, two lower unit cells were allowed to relax while all other atoms in the supercell were fixed at their bulk-like and previously optimized positions, which are shown in Table \ref{bulk_zellen}. A vacuum spacer of 2 nm was used to separate the copies of the periodic structures in the direction perpendicular to the surface. The ABO$_3$ perovskite structures possess a strong anisotropy resulting in the AO (A = Pb, Ba) and TiO$_2$ layers alternating in the [001] direction. The (001) surface of a ABO$_3$ can be terminated by a AO or a TiO$_2$ layer. Recently, Eglitis and Vanderbilt \cite{Eglitis2007} have reported for the cubic structure of BTO and PTO that the TiO$_2$-terminated surface is more stable. Using the same approach \cite{Eglitis2007} we calculated the surface energy for the both terminations of ABO$_3$. The results are shown in Table \ref{polar_per_layer}. For each  perovskite,  its TiO$_2$  terminated  (001) surface  is energetically favourable, indeed. Thus, we consider the (001) surface of ABO$_3$ to be TiO$_2$ terminated in the following.

In polar PTO and BTO, the displacements of the Ti and O atoms occur along the $z$-axis so that $\bf P$ is considered to be directed along [001] direction as well. First, we must formally set up the direction of electric dipole in the unrelaxed supercell assuming, for instance, that O is always above the corresponding cations in each layer along [001], as given in Table \ref{bulk_zellen}. Then we can model the two distinct situation alternatively placing the bulk-like 3-unit-cell thick substrate and relaxed layers against each other along the $z$ axis. In the first case, which we denote as $P_\downarrow$, the direction of $P$ is antiparallel to the surface normal. The second model labelled by $P_\uparrow$ corresponds to the case where all cations are above O before relaxation and, therefore, $P$ is parallel to [001]. In the tetragonal FE structure, both configurations may coexist in the random state as  $P_\downarrow$ and $P_\uparrow$ domains separated by a domain wall of $<$2 nm. To quantify the process of relaxation we use the cation-anion displacements $\delta  = z_{O} -z_{cation}$ calculated for each AO and TiO layer near the interface. In the bulk-like substrate of polar PTO and BTO, the model $P_\uparrow$ means that $\delta <0$ and {\it vice versa} the case of $P_\downarrow$ models the situation where $\delta >0$.

\begin{figure}
\includegraphics[width=0.9\linewidth]{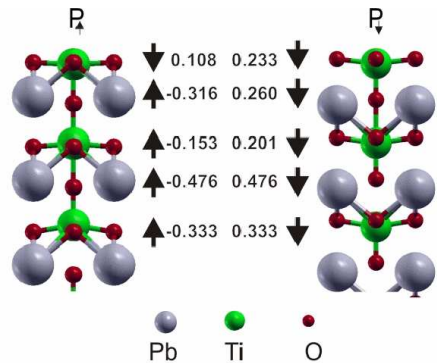}
\caption{\label{free_surface}The relaxed top layers of the PTO (001) surface, with polarization
 $P_\uparrow$ (left) and $P_\downarrow$ (right). The arrows indicate the direction of the dipole within each layer while the numbers show the displacements $\delta $ in \AA \, between O and Ti/Pb along [001]. For the case of $P_\uparrow$ the sign of $\delta $ in the topmost TiO layer is reversed.}
\end{figure}
Fig. \ref{free_surface} shows several top monolayers (ML) of PTO(001) after relaxation. The case $P_\uparrow$ ($P_\downarrow$) is shown on the left (right) side of Fig.\ref{free_surface}. The arrows indicate the direction of the dipoles in each ML, while the numbers at the arrows give the intralayer displacements $\delta$ in \AA \,, calculated between the O and metal atoms along [001]. In the state $P_\downarrow$, all dipoles possess the same orientation that means that O is always above the cation within each layer. In bulk PTO, the intralayer displacement, $\delta$, is 0.333\AA~for the TiO layer and 0.476\AA~for the PbO layer. For the three top ML near the interface, their $\delta$ are reduced by 30--45\% with respect to the corresponding bulk values. In the topmost TiO layer, the reduction of $\delta$ is $\sim$30\%. For the second (PbO) and third (TiO) layers from the interface, we find that their $\delta$ is reduced by 45\% and 40\%, respectively. In the case of $P_\uparrow$, shown in the left panel of Fig. \ref{free_surface}, the result of relaxation is rather different. In the third ML, the separation between Ti and O along [001] is 0.153 \AA~, which is reduced by 54\% against the corresponding bulk value. For the second ML, we obtain the reduction of 33\%. However, the most significant changes occur in the topmost TiO layer, whose $\delta$ is largely reduced by 68\% whereas the dipole is reversed compared to all others. Thus, using the $P_\uparrow$ model and placing all O below the cations, we obtained in the topmost layer the relaxed configuration where O is above Ti. This is similar to the case of $P_\downarrow$. 

\begin{figure}
\includegraphics[width=0.9\linewidth]{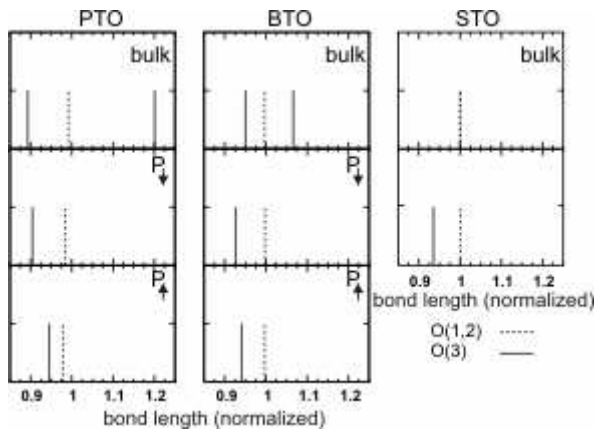}
\caption{\label{bonds} Ti-O bond lengths in the fivefold coordinated polyhedra of the topmost layer of PTO(001), BTO(001) and STO(001) compared to those obtained in the corresponding bulk structures, with the sixfold coordinated environments.}
\end{figure}

To investigate the effect of the surface rumpling in perovskites we repeat the calculations for STO(001) and BTO(001). In Table \ref{polar_per_layer} the corresponding results of our zero temperature calculation are listed. For paraelectric STO, we obtain that its TiO-terminated (001) surface after relaxation becomes marginally polar, with a positive rumpling normal to the surface in the top three ML where the O atoms are above the cations by $<$0.12 \AA~. This is in good agreement with the most recent experimental studies.\cite{Heide2001} The positive rumpling predicted for bare surfaces of perovskites leads to relatively low catalytic activity. With increasing temperature, the rumpling is distorted and it may stimulate further potential catalysis. For the TiO-terminated BTO surface, we have found the details of relaxation similar to those of PTO. In fact, all our results are in good agreement with those reported by Eglitis and Vanderbilt.\cite{Eglitis2007} In case of $P_\uparrow$, the topmost BTO rumpling of $\sim$0.1 \AA~ being larger than the corresponding bulk value, is similar to that of highly polar PTO. The sign of $\delta$ in the topmost ML of BTO is reversed with respect to all others calculated for the layers situated far down from the interface. In the third ML $\delta$ is 0. The $P_\downarrow$ model yields for BTO(001) the reversal dipole in the second ML, with marginal $\delta$. Thus, we find for the three systems and different arrangements of $\bf P$ that the TiO-terminated (001) surfaces prefer the configuration where O is above Ti.   
\begin{table}
\caption{\label{polar_per_layer} The surface energy E$_{surf}$ (in eV) for the TiO$_2$ and AO terminated (001)
surfaces and the cation-anion displacements $\delta$ (in \AA) calculated for the top five layers of the perovskite (001) surfaces.}
\begin{ruledtabular}
\begin{tabular}{cccccc}
  &  \multicolumn{2}{c}{PTO} & \multicolumn{2}{c}{BTO} & STO \\  
& $P_\uparrow$ &$P_\downarrow$&$P_\uparrow$&$P_\downarrow$&P=0\\
\hline
E$_{surf}$(AO)		&2.21&2.46&1.49&1.63&1.36\\
E$_{surf}$(TiO$_2$)	&2.07&2.20&1.31&1.32&1.34\\
\hline
ML& &&\\
1: O-Ti&0.108&0.233&0.102&0.129&0.072\\
2: O-A&-0.316&0.260&-0.086&-0.030&0.118\\
3: O-Ti&-0.153&0.201&-0.002&0.067&0.018\\
4: O-A&-0.476&0.476&-0.082&0.082&0.0\\
5: O-Ti&-0.333&0.333&-0.086&0.086&0.0
\end{tabular}
\end{ruledtabular}
\end{table}
In the cubic ABO$_3$ perovskite structure, each Ti$^{4+}$ ion sits in the regular sixfold coordinated site with all of the Ti-O bonds of equal length, as shown in Fig. \ref{bonds} for bulk STO. 

In the tetragonal perovskite structure, such as $t$-PTO, the relaxed cluster of O atoms about the sixfold coordinated Ti forms a distorted octahedron, where one of the two bond lengths along [001] is rather short while another Ti-O bond in the vertical direction is significantly longer than the four other bonds stretching in the equatorial plane. If we exclude the longest Ti-O bond from the consideration using the electrostatic arguments then the environment for each Ti becomes the fivefold coordinated polyhedra, which is similar to that of Ti at the interface. The left three panels of  Fig. \ref{bonds} compare the Ti-O bond lengths in bulk PTO, normalized to the value of the ideal octahedron in the cubic structure, to those in the polyhedron around fivefold coordinated Ti in the topmost layer. Using the $P_\downarrow$ model for the PTO (001) surface, we obtain the Ti-O bonds whose lengths are similar to those of $t$-PTO and, hence, nothing dramatic happens in the environment of the topmost Ti. In the case of $P_\uparrow$, the bond length distribution around the topmost Ti is restricted so that the equatorial and vertical bond lengths tend to be equal to each other. Moreover, the closest O atom to the surface Ti is attached along [001] from the opposite side compared to all Ti placed below the surface in the regular crystal structure within $P_\uparrow$. It is clear that the O-Ti-O bond angles for the equatorial Ti-O bonds of the topmost ML must be dramatically changed to compensate the charge distribution around Ti. It appears that these bond angles become $>$90$^{\circ }$, as shown in Fig. \ref{free_surface}. Therefore, whatever the state of ${\bf P}$ is modeled in $t$-PTO, the O atoms must relax above Ti on the Ti-O terminated (001) surface. Regarding BTO, the same conclusions may be drawn.   
  
\begin{figure}[b]
\includegraphics[width=0.6\linewidth]{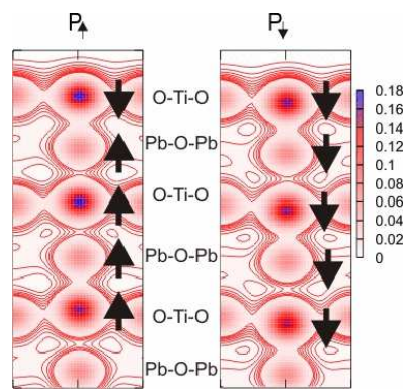}
\caption{\label{charge_pbtio} The charge density, calculated for the top six layers of PbTiO$_3$(001) and projected on the x-z plane. The two opposite polarization states are shown: $P_\uparrow$ (left) and $P_\downarrow$ (right) where the dipole directions within each layer is labelled by the arrows. In the case of $P_\uparrow$, the isocharge lines of the third TiO layer from the interface show a blockade of the charge transfer along [001].}  
\end{figure}

Fig. \ref{charge_pbtio} shows the view of the charge density along [010]  calculated for the top six ML of PTO and projected on the x-z plane of the supercell.
The isocharge lines plotted in Fig. \ref{charge_pbtio} for both cases of ${\bf P}$ illustrate the charge transfer across the cell while the arrows indicate the dipole directions within each ML. In the case of $P_\downarrow$, which is shown in the right panel of Fig. \ref{charge_pbtio}, there are three bridges seen between Ti and nearest O. The shortest bond with O, which is always below Ti along [001], has the large population value. In the $P_\uparrow$ state, the charge transfer picture is similar to that of $P_\downarrow$ for the topmost Ti only. Far below the interface (starting from the 5th ML) all Ti have the most populated bond with O which is above Ti. In the third ML, however, the Ti ion is strongly bonded to equatorial oxygens showing some sort of blockade for the charge transfer along the [001] direction. This may reveal the key electronic states factors behind the surface relaxation of polar FE. 

Recently, Urakami et al.\cite{Urakami2007} have observed the surface conductance of BaTiO$_3$ single crystals in ultra high vacuum below $T_C$. It has been shown that the in-plane conductance is the result of an intrinsic surface electron/hole layer that is, due to the surface polarity and not due to O vacancies or some other defects. The $I-V$ characteristics shows a pronounced difference of conduction between the poled states in BTO. We can explain this difference in a simple way using our {\it ab initio} results.       
\begin{figure}
\includegraphics[width=0.8\linewidth]{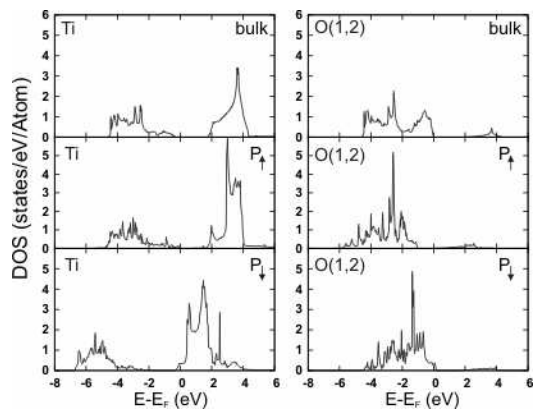}
\caption{\label{bto_dos}
The Ti- and O$_1$-resolved DOS of $t$-BTO are plotted in the two upper panels where the top of the valence band is taken as zero energy. In the lower panels, the corresponding local DOS of the TiO-terminated (001) surface are plotted for the $P_\uparrow$ and $P_\downarrow$ states.}   
\end{figure}
To reveal the differences between $P_\downarrow$ and $P_\uparrow$ we plot in Fig. \ref{bto_dos} the Mulliken site-projected density of states (DOS) of the BTO(001) surface for both cases of polarization. The Ti and O DOS for the topmost ML are shown in comparison with the corresponding DOS of $t$-BTO. For bulk BTO, Fig. \ref{bto_dos} shows a pronounced insulating band gap of 2 eV. The value is typically underestimated by the LDA approximation of DFT. Comparing the Ti and O DOS of $t$-BTO and the topmost ML of BTO(001), we see a spectacular change of the electronic states occurring due to the surface relaxation and variation of ${\bf P}$. The major DOS features can be summarized as follows. In the case of $P_\uparrow$, a few O states appear in the band gap while the Ti DOS is not affected. For the $P_\downarrow$ poled state, the Ti lower conduction band, being shifted downwards in energy by $\sim$2 eV, significantly contributes to the DOS in the band gap region. This causes metallic behavior of the topmost ML in the case of $P_\downarrow$, yielding rather large in-plane conductance. In the case of $P_\uparrow$ the Ti states have a gap at $E_f$ which is related to a tiny in-plane conductance. Depending on the polarization direction the topmost ML undergoes a transition from metallic to oxide behavior  shows metallic or oxide behavior. As a consequence the in-plane conductance changes drastically which is a reasonable explanation of the experimental results by Urakami et al.\cite{Urakami2007}.

In summary, from the {\it ab initio} basis of our work we have shown that the intrinsic ferroelectricity in polar perovskites is suppressed by $\sim$30\% in the surface region. For both cases of polarization direction, the TiO terminated surface of BTO and PTO forms an electric dipole where the O atoms being shifted $>$0.1 \AA~ above Ti. But nevertheless the electronic structure of the surface layer changes from metallic to oxide behavior under reversal of polarization which changes the surface conductance drastically.\cite{Urakami2007} This may have important implications in the design of multiferroic nano-devices.   



\end{document}